\documentclass[useAMS,usenatbib,letterpaper]{mn2e}

\usepackage{ulem}
\usepackage{amsmath}
\usepackage{graphicx}
\usepackage{fixltx2e}
\usepackage{placeins}
\usepackage{multirow}
\usepackage{tabulary}
\usepackage[para*]{manyfoot}
\DeclareNewFootnote[para]{B}[symbol]

\newcommand{\hii}{\mbox{H\,{\sc ii }\,}}

\newcommand{\mjy}{\mbox{mJy\,beam$^{-1}$}}

\newcommand{\footnoteremember}[2]{\footnote{#2} \newcounter{#1} \setcounter{#1}{\value{footnote}}}

\title[A multi-wavelength study of the radio source G296.7-0.9]{A multi-wavelength study of the radio source G296.7-0.9: confirmation as a Galactic supernova remnant}
\author[W.J. Robbins et al.]{W.J. Robbins$^{1,}$\thanks{E-mail: wrobbins@physics.usyd.edu.au}, B.M. Gaensler$^{1,2}$, T. Murphy$^{1,3}$, S. Reeves$^{1}$, and A.J. Green$^{1}$ \\
$^{1}$Sydney Institute for Astronomy, School of Physics, The University of Sydney, NSW 2006, Australia\\
$^{2}$Australian Laureate Fellow\\
$^{3}$School of Information Technologies, The University of Sydney, NSW 2006, Australia}

\begin{document}

\date{Accepted 2011 September 27. Received 2011 September 2; in original form June 8}

\pagerange{\pageref{firstpage}--\pageref{lastpage}} \pubyear{2011}

\maketitle

\label{firstpage}

\begin{abstract}
We present a multi-wavelength study of the radio source G296.7-0.9. This source has a bilateral radio morphology, a radio spectral index of $-0.5 \pm 0.1$, sparse patches of linear polarisation, and thermal X-rays with a bright arc near the radio boundary. Considering these characteristics, we conclude that G296.7-0.9 is a supernova remnant (SNR). The age and morphology of the SNR in the context of its environment suggest that the source is co-located with an \hii region, and that portions of the shock front have broken out into a lower density medium. We see no evidence for a neutron star or pulsar wind nebula associated with SNR G296.7-0.9. 
\end{abstract}


\section{Introduction}
Supernovae (SNe) and their remnants are significant drivers of galactic evolution \citep{tammann:1994,fitzpatrick:2004,nomoto:2006}. The details of the explosion and properties of the local environment combine to determine the morphology of a supernova remnant (SNR). Developing a more general understanding of SN explosions, the evolution of the subsequent shocks, and the effect on their galaxy requires a statistically significant sample of objects for which the explosive and environmental effects have been decoupled.
\par
Here we present one such case study: a multi-wavelength analysis of the SNR candidate G296.7-0.9 and its environment. This source was initially classified as \hii region G296.593-0.975 based upon its association with radio recombination lines and an apparently thermal spectrum \citep[and references therein]{manchester:1969,caswell:1987,kuchar:1997}. However, the ROSAT all-sky X-ray survey \citep{voges:1999} detected emission from this source, which prompted the suggestion that it might be part of a supernova remnant centred at G296.7-0.9 \citep{schaudel:2002a,schaudel:2002b,gamarova:2002}. 
\par
In section \ref{analysis}, we present an analysis of archival radio, infrared, optical, and X-ray data of G296.7-0.9, which confirm it as a SNR. We discuss the source's environment, properties, and morphology in section \ref{discussion}.


\section{Observations and Results}\label{analysis}
\subsection{Radio}\label{morph:anal}
\subsubsection{Australia Telescope Compact Array}
The Australia Telescope Compact Array (ATCA) is a radio interferometer consisting of six antennas, each 22 m in diameter, located near Narrabri, New South Wales, Australia. Five of these antennas are usually on a 3 km east-west track. To provide a maximum possible baseline of 6 km, a sixth antenna is fixed to a pad at a distance of 3 km from the end of the 3 km east-west track. 
\par
We extracted ATCA data from the Australia Telescope Online Archive\footnote{\scriptsize \url{http://atoa.atnf.csiro.au}} for observations of G296.7-0.9 made on 2001 Sep~1 and 2002 Jan~10. These are continuum observations with data collected at a central frequency of 1384 MHz in each of the 6B (2001 Sep) and 750A (2002 Jan) array configurations.  Both observations include twelve 8 MHz, full-polarisation channels around the central frequency. The primary beam size is roughly 33 arcmin (FWHP).\footnote{\scriptsize\url{http://www.atnf.csiro.au/observers/docs/ca\_obs\_guide/ca\_obs\_guide.pdf}} Flux calibration was performed relative to PKS B1934-638, which has a flux density of 14.94 $\pm$ 0.1 Jy at 1384 MHz. Phase calibration and polarisation leakage were determined twice per hour by observations of PKS B1148-671. 
\par 
The MIRIAD package\footnote{\scriptsize\url{http://www.atnf.csiro.au/computing/software/miriad}} was used in all stages of ATCA data reduction. To form an image, we used data from both the 750A and 6B configurations, but excluded baselines with u-v distances larger than 4 k$\lambda$. This upper-limit on u-v distance was chosen to eliminate gaps in u-v coverage and provide the best signal to noise ratio. The image shown in Figs.~\ref{fig.radio}(a) and \ref{fig.radio}(b) is a 1384 MHz image of G296.7-0.9, created using uniform weighting and a maximum entropy deconvolution method. This image has not been corrected for the attenuation of the primary beam away from the phase centre. The resolution is 45 x 33 arcsec$^2$ and the noise level, $\sigma$, is 0.46 \mjy .  The labels in Fig.~\ref{fig.radio}(a) are the only features that distinguish it from Fig.~\ref{fig.radio}(b).
\par
The following features are labelled in Fig.~\ref{fig.radio}(a). G296.7-0.9 has two high surface brightness limbs that are split by an axis oriented 140 $\pm$ 20 deg east of north. Emission along this axis is markedly reduced, save for three narrow elongated structures (NESs) that appear to connect the east and west limbs. There is a bulge-like enhancement of emission (``bulge") in the western limb with a peak surface brightness at about 1.5 times the peak in the eastern limb. Using the $3\sigma$ level of Fig.~\ref{fig.radio}(b) to define the outer boundary of the source, it is clear that there are two low surface brightness arms which extend from the eastern lobe to the north and the southeast, plus a plateau of emission extending southwest from the western lobe. Two depressions of surface brightness exist between the western limb and plateau and are roughly bounded by the 10$\sigma$ level.
\par
We define the source centre as the average of the midpoints of circles fit to the major and minor axes of G296.7-0.9, at $\alpha_{2000}$=11:55:34.69, $\delta_{2000}=-$63:06:42. We define a reduced diameter in terms of the length of the major and minor axes, $\sqrt{\rm{major \times minor}}=11 \pm 2$ arcmin. (The large fractional error is related to the uncertainty in fitting a circle to the irregular boundary of the source.) After correcting for the attenuation of the primary beam, the flux density of G296.7-0.9 is 2.5 $\pm$ 0.2 Jy at 1.4 GHz.
\par
There are several unrelated sources in the field; these are numerically labelled in Fig.~\ref{fig.radio}(a).  Source `1' is the planetary nebula candidate G296.8-0.9 \citep{parker:2006,cohen:2007,ramos-larios:2008,kwok:2008} and is located southeast of G296.7-0.9, near a second unresolved source labelled `2'. A diffuse patch of radio emission with a bright central core (possibly a background source) is  labelled `3'.
\subsubsection{Molonglo Observatory Synthesis Telescope}
The second epoch Molonglo Galactic Plane Survey \\ \citep[MGPS-2; ][]{murphy:2007} was conducted at a frequency of 843 MHz over the southern Galactic plane using the Molonglo Observatory Synthesis Telescope (MOST). The images have a resolution of 43 $\times$ 43 $\csc{|\delta|}$ arcsec$^2$ and a sensitivity level of $\sim$ 1 mJy beam$^{-1}$. Because of its lower frequency and continuous baseline coverage from 13 to 1600~m, synthesis images from MOST are sensitive to larger angular scales than the ATCA data. We acquired an image of the field including G296.7-0.9 from the MGPS-2 archive\footnote{\scriptsize\url{http://www.physics.usyd.edu.au/sifa/Main/MGPS2}}. 
\par
\begin{figure*}
\centering
\includegraphics[scale=.5,angle=-90]{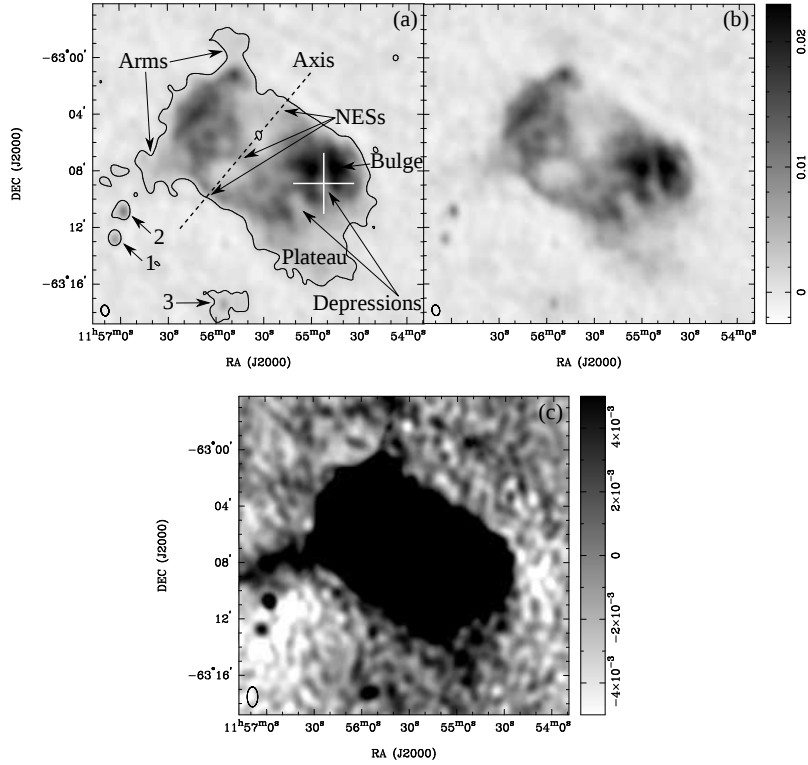}
\caption{Radio images of G296.7-0.9. Panels (a) and (b) are identical 1.4 GHz ATCA images, apart from the labels in panel (a). The scale is given in Jy beam$^{-1}$ and the 45 x 33 arcsec$^{2}$ synthesised beam is represented by the open ellipse shown in the bottom left of both panels. In panel (a), sources unrelated to G296.7-0.9 are labelled with numbers, and the pointing centre and 4.4 arcmin beam size of the radio recombination line measurement reported by \protect \cite{caswell:1987} are represented by a white `+'. The contour is the $3\sigma$ level of the image, 1.38 \mjy . Panel (c) is a 843 MHz MOST image of G296.7-0.9 with a 43 x 49 arcsec$^{2}$ synthesised beam, illustrated by the open ellipse in the lower left \protect \citep{murphy:2007}. The saturated Jy beam$^{-1}$ scale has been chosen to highlight the low surface brightness `arms' and `plateau.'}
\label{fig.radio}
\end{figure*}

The morphology of G296.7-0.9 at 843 MHz is very similar to the 1.4GHz images presented in Figs.~\ref{fig.radio}(a) and \ref{fig.radio}(b). For this reason, we have chosen a saturated scale for Fig.~\ref{fig.radio}(c), a cutout from the MGPS-2 of the G296.7-0.9 field, in order to highlight that the `arms' and `plateau' are similar in the ATCA and MOST images. The resolution of Fig.~\ref{fig.radio}(c) is 43 $\times$ 49 arcsec$^2$, the peak surface brightness is 42.6 \mjy , and the noise level is 1.6 \mjy .

\subsubsection{Radio Spectral Index}
We have used the images in Figs.~\ref{fig.radio}(b) (ATCA) and \ref{fig.radio}(c) (MGPS-2) to determine a spectral index for G296.7-0.9. The six dishes of the ATCA sparsely sample the u-v plane, so the images presented in Figs.~\ref{fig.radio}(b) and \ref{fig.radio}(c) necessarily measure flux at a different range of spatial scales. Thus, a direct comparison of fluxes from the two images could yield a spectral index which has too little flux at high frequencies.  
\par
We have resampled the MGPS-2 image using the less complete u-v spacings of the ATCA data \citep[cf.][]{gaensler:2003}. This has been accomplished by creating a sky model of the source as seen by MOST, then sampling this sky model with the composite 750A-6B synthesised ATCA beam. Thereby, we have introduced the ATCA u-v sampling to the brightness distribution as measured in the MGPS-2. 
\par
To look for spatial variations in the spectral index, we have used spectral index tomography \citep{katz-stone:1997}, which assumes the flux is a power-law in frequency, S$_\nu \propto \nu^{\alpha}$. Given two images, $i$ and $j$, which are well-spaced in frequency, we iterated through possible values of spectral index and made an image of the residual, R~$= I_i - I_j \left(\nu_i /\nu_i \right)^\alpha$. The residual  vanishes when $\alpha$ is equivalent to the spectral index of emission, so sections of the map with different spectral indices are readily identifiable. 
\par
\begin{figure*}
\centering
\includegraphics[scale=0.4,angle=-90]{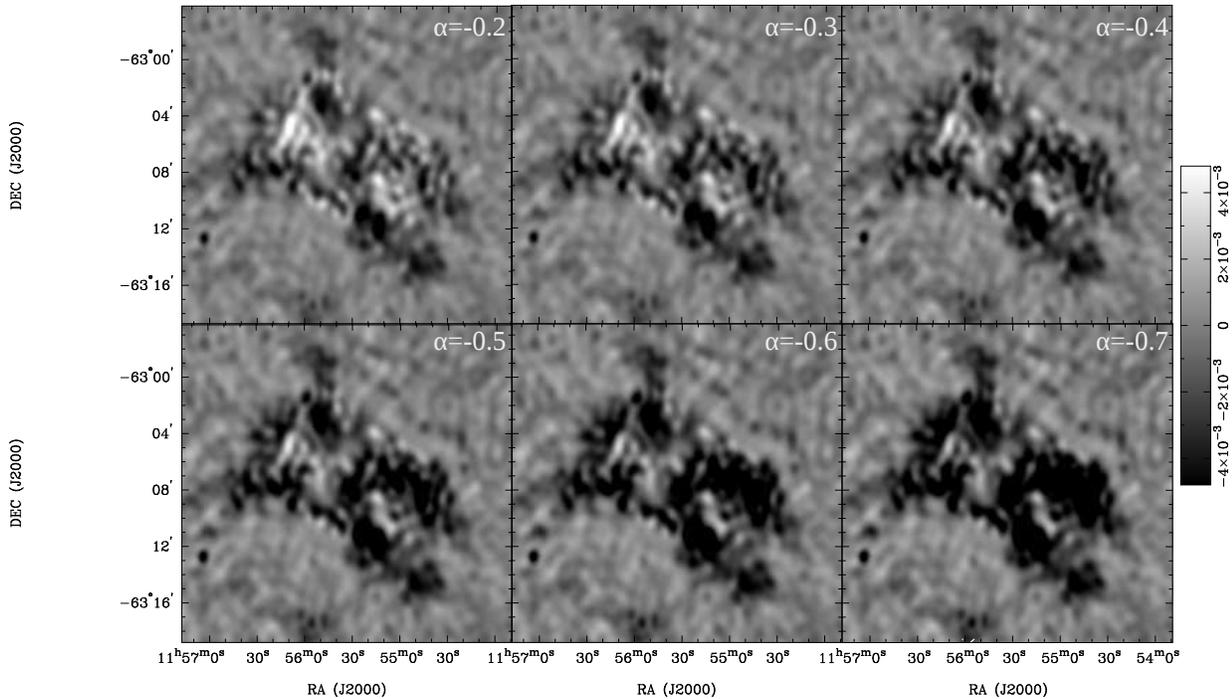}
\caption{Radio spectral index tomography images of G296.7-0.9. These images are the residual of total intensity, $R=I_{843}-I_{1384}\left(\nu_{843}/\nu_{1384}\right)^\alpha$, between the 843 and 1384 MHz images presented in Fig. \ref{fig.radio}, for an assumed spectral index of $\alpha$. The spectral index for each residual image is labelled in the upper-right corner of each panel and the wedge is given in Jy beam$^{-1}$.}
\label{alpha}
\end{figure*}

Fig.~\ref{alpha} displays six residual images for spectral index choices of -0.2 to -0.7. We find that well-detected portions of G296.7-0.9 have a uniform spectral index of $ \alpha = -0.5 \pm 0.1 $.

\subsubsection{Polarisation}\label{anal:pol}
To mitigate bandwidth depolarisation, we have formed separate linear polarisation images for each of the twelve ATCA channels, L = $\sqrt{\mathrm{Q}^2+\mathrm{U}^2-\eta^2}$, where the $\eta$-term corrects the Ricean bias that results from rectifying Gaussian noise in the Q- and U-images. 
\par
\begin{figure}
\begin{center}
\includegraphics[width=0.35\textwidth,angle=-90]{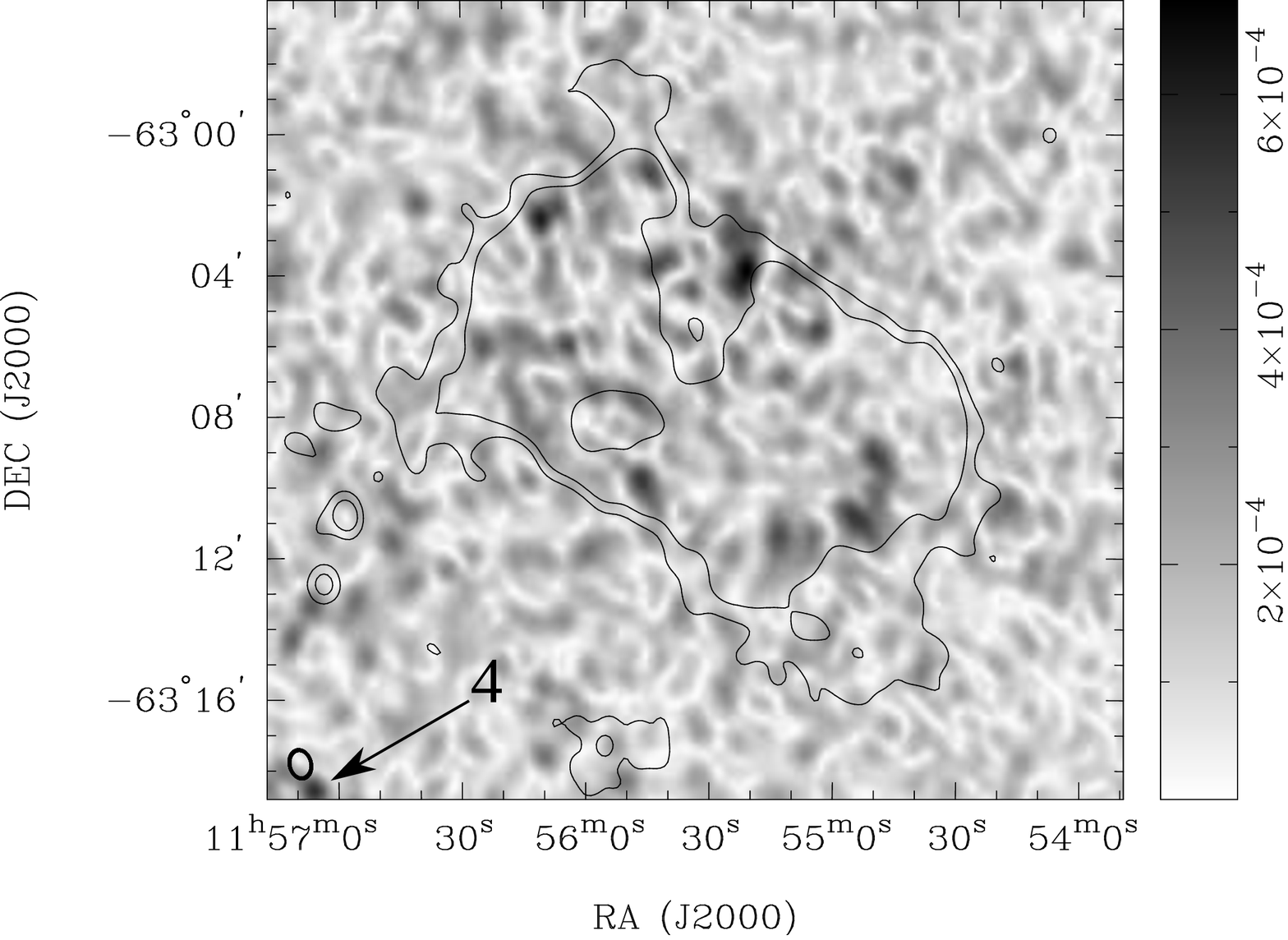}
\caption{Linear polarisation detection image of G296.7-0.9 at 1.4GHz. This image is the error-weighted average of the Ricean-biased ($\eta$=0) images of linear polarisation ($L=\sqrt{Q^2+U^2-\eta^2}$) from the 12 ATCA channels. The contours are the 3 and 10$\sigma$ radio contours of Fig.~\ref{fig.radio}(b). The source labelled `4' is unrelated.}\label{pol}
\end{center}
\end{figure}

A polarisation detection image ($\eta$=0) is shown in Fig. \ref{pol}. We observe several marginally polarised pockets of emission, largely around the edges of the source, which are polarised at approximately 5\% of total intensity. Using rotation measure synthesis \citep{burn:1966,brentjens:2005,heald:2009}, we find rotation measures for these patches of between +10 and +260 rad m$^{-2}$, with an average of +155~$\pm$~53~(statistical)~$\pm$~27~(systematic)~rad~m$^{-2}$.

\subsection{X-ray}
\subsubsection{ROSAT}
A pair of redundant Position Sensitive Proportional Counters (PSPCs) were mounted on the X-ray Telescope launched on the R\"{o}entgen Satellite (ROSAT). These detectors were used to conduct an all-sky survey in the 0.1 to 2.4 keV energy range \citep{voges:1999}. The resulting data have a 20 arcsec on-axis spatial resolution and energy resolution of 0.41 keV at 1 keV, which degrades as the square-root of energy. 
\par
\begin{figure}
\begin{center}
\includegraphics[width=0.35\textwidth,angle=-90]{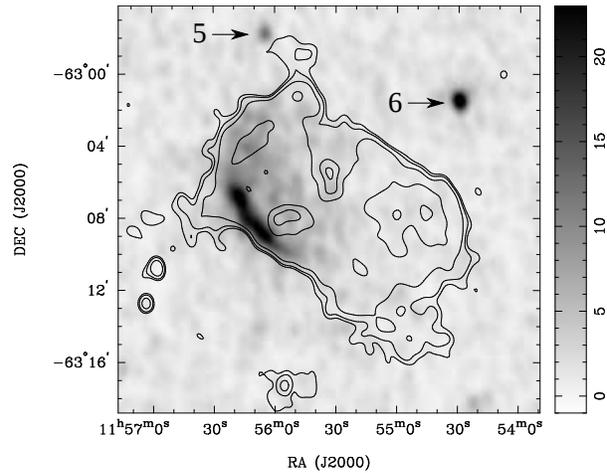}
\caption{Combined 9.0 ks ROSAT PSPC image of G296.7-0.9 in the 0.1 - 2.4 keV energy range. The image has been smoothed to the resolution of the 1.4 GHz image in Fig.~\ref{fig.radio}(b) and overlaid with its 3, 5, 10, 30, and 50$\sigma$ contours. The units are counts beam$^{-1}$. Two unrelated X-ray sources are numbered `5' and `6'.}\label{pspc}
\end{center}
\end{figure}

 
G296.7-0.9 was observed twice with a PSPC detector, once on 1993 Feb 1--4 and again on 1998 Feb 19--21, for 4.5 and 4.7 ks, respectively. We retrieved data for these observations from the W3Browse service\footnote{\scriptsize\url{http://heasarc.gsfc.nasa.gov/W3Browse}}. Fig.~\ref{pspc} is a cutout image from the composite ROSAT PSPC dataset that has been smoothed to the resolution of Fig.~\ref{fig.radio}(b); the contours are the 3, 5, 10, 30, and 50$\sigma$ contours of the radio emission in Fig.~\ref{fig.radio}(b).
\par
There is a bright X-ray arc in the southeast of G296.7-0.9 that bounds low levels of diffuse X-ray emission throughout the eastern half of the source. Except for two unrelated sources marked `5' and `6' in Fig.~\ref{pspc}, the X-ray emission from G296.7-0.9 is roughly bounded by the 10$\sigma$ radio contour.
\par
We used the FTOOLS available in the Standard Hera service\footnoteremember{hera}{\scriptsize\url{http://heasarc.gsfc.nasa.gov/hera}}(HERA) to combine event files and prepare spectra. An ancillary response file, which contains calibration information for the effective area and quantum efficiency of the PSPC detectors as a function of energy, was created using the pcarf FTOOL. The redistribution matrix file, pspcb\_gain2\_256.rmf, which allows for the conversion of spectral channels to energy bins for on- and off-axis observations taken after 1991 Oct 14, was obtained from the ROSAT FTP area\footnote{\scriptsize\url{ftp://legacy.gsfc.nasa.gov/caldb/data/rosat/pspc/cpf/matrices}}.
\par 
We have extracted a source spectrum from a circular region of Fig.~\ref{pspc} with a radius of 5 arcmin, centred at $\alpha_{2000}$= 11:55:46, $\delta_{2000}$= -63:05:41. The background spectrum was taken from an annulus with an inner radius of 5 arcmin and an outer radius of 20 arcmin, concentric to the source region.
\par
The spectra were grouped with a minimum of 20 photons per energy bin and spectral fitting was performed with version 11 of XSPEC \citep{arnaud:1996}. Bad spectral channels were ignored during fitting, as was the lowest energy bin because the higher than expected number of counts in this channel shift the peak of our fits toward unrealistically low energies.
\par
We have fit the following simple models, which are described in detail in the XSPEC manual\footnote{\scriptsize \url{http://heasarc.gsfc.nasa.gov/docs/xanadu/xspec/manual/XspecModels.html}}. Assuming solar ion abundances and photoelectric absorption along the line-of-sight, we have fit the bremsstrahlung, MEKAL, Raymond-Smith, non-equilibrium ionisation (NEI), and power-law models to our data.
\par
The fit statistics, parameters, and fluxes for these models are presented in Table~\ref{x-fits-table}. For all models, the $\chi^2$ and number of degrees of freedom (DOF) are given as a quotient and the integrated column density of absorbing particles is given in the n$_\mathrm{H}$ column. Thermal models include a parameter for the X-ray temperature (kT) in units of energy, whereas the power-law model's photon index is represented by the dimensionless $\Gamma$ parameter in the temperature column. The NEI model includes a parameter that quantifies the equilibration timescale, $\tau$, as the product of the electron density and time to reach ionisation equilibrium; that is, for a fixed $\tau$, an increase in electron density results in quicker thermalisation. 
\par
Based upon the reduced-$\chi^2$ of our fits, we conclude that the dominant mechanism in our source region is thermal. Though our data does not allow us to favour any one model, an absorbed MEKAL model is formally the best-fit. The MEKAL model approximates the combined bremsstrahlung continuum and numerous recombination lines of multiply-ionised elements that are expected from a hot, optically-thin plasma in thermal equilibrium. 
\par
Note that because of the low number of counts, we were not able to fit a spectrum to the bright X-ray arc in the southeast of G296.7-0.9 without including the low level emission throughout the rest of the eastern lobe.

\begin{small}
\begin{table*}
 \begin{minipage}{220mm}
\caption{ROSAT emission from G296.7-0.9: emission models, fit statistics, parameters with 1$\sigma$ errors, and 0.5 - 2.4 keV fluxes}\label{x-fits-table}
\begin{tabular}{ccccccc}
\hline Model & $\chi^2$/DOF & kT & n$_{\mathrm{H}}$ &  $\tau$ & Absorbed Flux & Unabsorbed Flux\\
& & (eV)& (10$^{21}$ cm$^{-2})$& (cm$^{-3}$ s$^{-1}$)&\multicolumn{2}{c}{($10^{-12}$ ergs s$^{-1}$ cm$^{-2}$)}  \\
\hline 
\vspace{2mm}  MEKAL & 58.6/67 & 110$^{+2}_{-4}$ & 5.3$\pm 0.6$& &$\sim$2 & $\sim$100 \\

\vspace{2mm} Raymond-Smith & 59.6/67 & 116$^{+2}_{-2}$ & 4.7$^{+0.9}_{-0.6}$ &  &$\sim$2 & $\sim$70\\

\vspace{2mm} Non-Equilibrium Ionisation & 59.3/66 & 200$^{+200}_{-100}$ &  3$^{+1}_{-2}$& 2$^{+3}_{-1}\times10^{10}$ &$\sim$2& $\sim$30 \\

\vspace{2mm} Bremsstrahlung & 66.0/67 & 107$^{+10}_{-7}$ & 4.7$^{+0.8}_{-0.7}$ &  &$\sim$2 &  $\sim$40\\

\vspace{2mm} Power-Law & 102.7/67 & $\Gamma$=9$^{+0.4}_{-0.4}$ & 6.6$\pm0.4$ & &$\sim$2& $\sim$100 \\

\hline

\end{tabular}
\end{minipage}
\end{table*}
\end{small}

\subsection{Infrared}
\subsubsection{Spitzer Space Telescope}
The Galactic Legacy Infrared Mid-plane Survey Extraordinaire \citep[GLIMPSE; ][]{churchwell:2009} was conducted with the Infrared Array Camera aboard the Spitzer Space Telescope (SST). Data were taken from each of four channels with central wavelengths of 3.6, 4.5, 5.8, and 8.0 $\mu$m and each has a resolution of 1.2 arcsec. Images from all four bands were obtained from the NASA/IPAC Infrared Science Archive\footnote{\scriptsize\url{http://irsa.ipac.caltech.edu/data/SPITZER/GLIMPSE}}. 

\begin{figure*}
\centering
\includegraphics[width=0.75\textwidth,angle=-90]{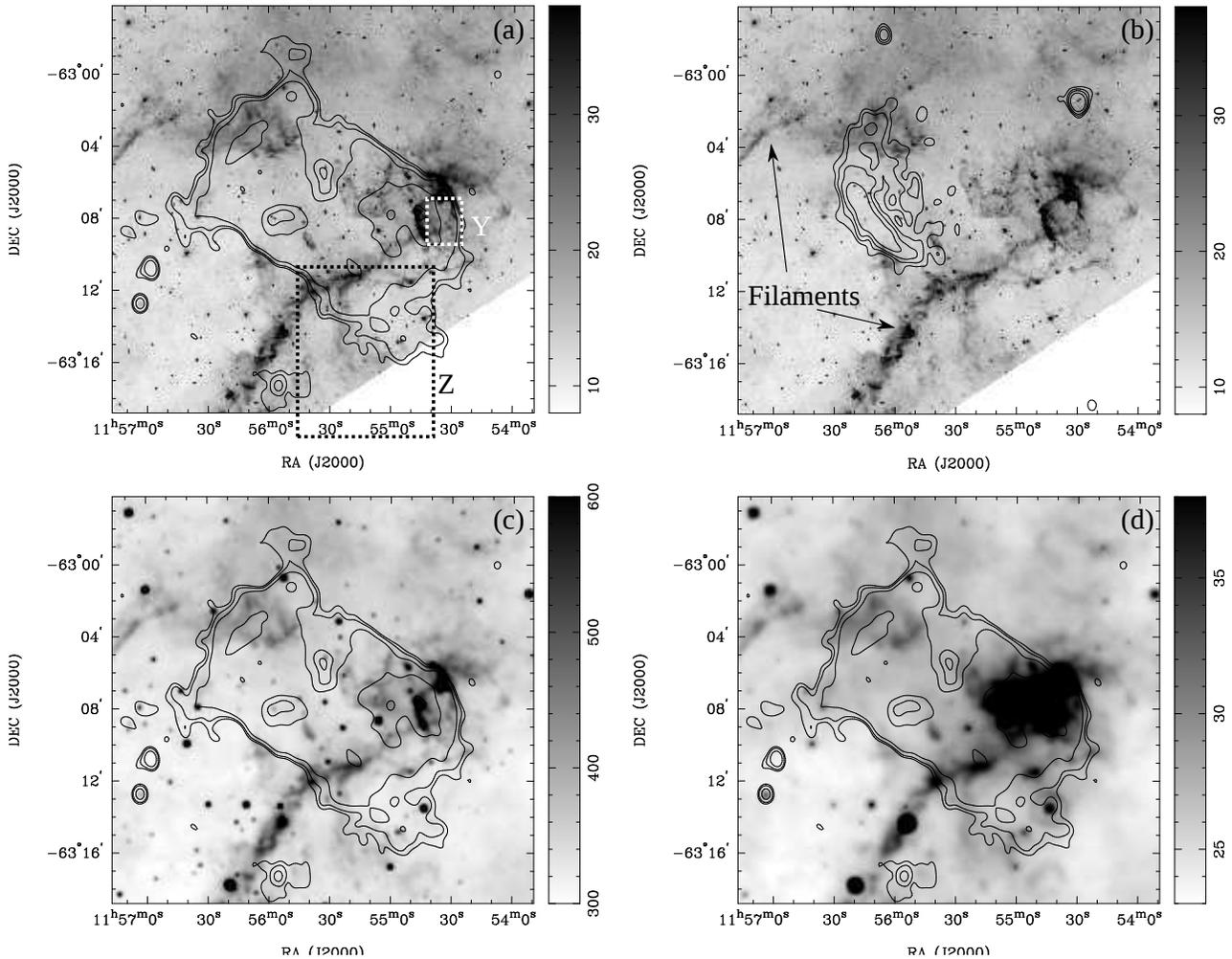}
\caption{Infrared images of G296.7-0.9. (a) GLIMPSE 8 $\mu$m image with 3, 5, 10, 30, and 50$\sigma$ levels of 1.4 GHz emission from Fig.~\ref{fig.radio}(b). The scale is in units of MJy sr$^{-1}$ and the resolution is 1.2 arcsec. The region labelled `Z' identifies the southwest portion of the plateau shown in Fig.~\ref{zoom.plateau} and `Y' is the portion of the western lobe displayed in Fig.~\ref{zoom.halpha}. (b) GLIMPSE 8 $\mu$m image where the contours are the 4.95, 6.75, 9.0, 13.5 counts beam$^{-1}$ levels from the X-ray image in Fig.~\ref{pspc}. The resolution of the image is 1.2 arcsec, the scale is in units of MJy sr$^{-1}$ and the `filaments' labelled in this figure are the $>20$ MJy sr$^{-1}$ features referred to in the text. (c) WISE 12 $\mu$m image where the levels are the 3, 5, 10, 30, and 50$\sigma$ contours of 1.4 GHz emission from Fig.~\ref{fig.radio}(b). The resolution is 6.5 arcsec and scale is in `data number' units described by \protect \cite{wright:2010}. (d) WISE 22 $\mu$m image where the levels are the 3, 5, 10, 30, and 50$\sigma$ contours of 1.4 GHz emission from Fig.~\ref{fig.radio}(b). The scale is in `data number' units described in \protect \cite{wright:2010} and the resolution is 12 arcsec.}
\label{fig.ir}
\end{figure*}
\par 
The GLIMPSE 8~$\mu$m image is displayed in Figs.~\ref{fig.ir}(a) and \ref{fig.ir}(b). The contours in Fig.~\ref{fig.ir}(b) are arbitrarily chosen X-ray levels of 4.95, 6.75, 9.0, 13.5 counts beam$^{-1}$ from Fig.~\ref{pspc}, whereas Fig.~\ref{fig.ir}(a) has the 3, 5, 10, 30, 50$\sigma$ levels of Fig.~\ref{fig.radio}(b). Most of the 8$\mu$m sources are unresolved, but are apparent in each of the GLIMPSE bands and are not detected in radio or X-rays. The 5.6 and 8 $\mu$m image additionally show high surface brightness filaments which are not identifiable at 3.6 and 4.5 $\mu$m. These filaments are most apparent at 8 $\mu$m, where their surface brightness is $>$ 20 MJy sr$^{-1}$. These are labelled in Fig.~\ref{fig.ir}(b).
\par
No area in the field is completely devoid of 8 $\mu$m emission. The lowest surface brightness emission ($\sim$ 10 MJy sr$^{-1}$) exists between the filaments and east of $\alpha_{2000}$=11:56:10, overlapping the southeastern arm, and in a second region that includes the plateau. Moderate levels ($\approx$ 15 MJy sr$^{-1}$) of emission cover the largest fraction of the field, north and west of G296.7-0.9. Both filaments cross some portion of the radio boundary of G296.7-0.9 and the southernmost trail extends past the northwestern edge of G296.7-0.9. 
\par 
Following \cite{reach:2006}, we normalise images from each GLIMPSE band to the surface brightness of the 8~$\mu$m image and find typical values for the filaments of 0.05, 0.07, 0.3 and 1 for central wavelengths of 3.6, 4.5, 5.8, and 8~$\mu$m, respectively. These values are consistent with expected contributions of polycyclic aromatic hydrocarbons (PAHs) \citep[see fig. 1 of][]{reach:2006}. We have used the method of \cite{cohen:2007}, essentially comparing the GLIMPSE 4.5, 5.8, and 8~$\mu$m bands, to examine the spatial distribution of PAHs. The strongest PAH emission occurs at $\alpha_{2000}$=11:54:46 and $\alpha_{2000}$=11:54:35, which are also the highest 8~$\mu$m surface brightness features in the field. We conclude that the filaments are dominated by PAHs in the GLIMPSE wavebands.
\par
\begin{figure}
\begin{center}
\includegraphics[width=0.45\textwidth]{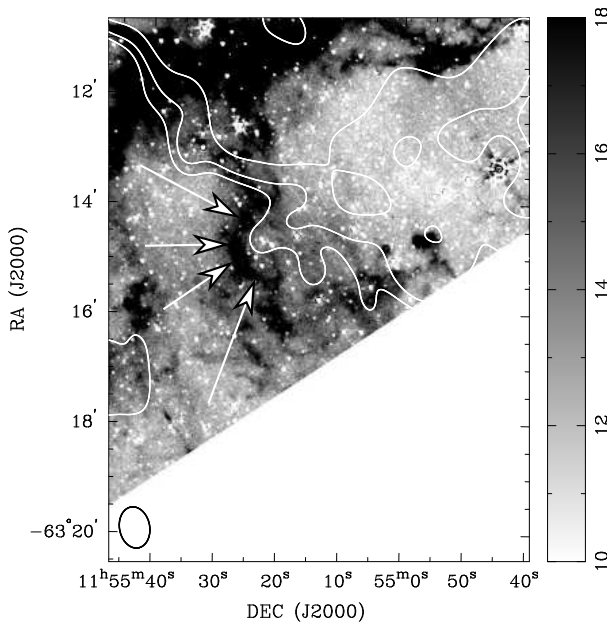}
\caption{Continuum-subtracted 8$\mu$m GLIMPSE image of the region labled `Z' in Fig.\ref{fig.ir}(a).  This image was formed by subtracting the 3.6~$\mu$m image (not shown) from the 8 $\mu$m image in order to isolate the strong 8 $\mu$m features. The scale is MJy sr$^{-1}$ and the resolution is 1.2 arcsec. The contours are the 3, 5, 10, 30, and 50$\sigma$ levels from Fig~\ref{fig.radio}(b) and the ellipse at the lower left is the resolution of these contours. The arrows highlight a region in which the 8 $\mu$m and radio emission may share morphological similarity.}\label{zoom.plateau}
\end{center}
\end{figure}

 
There may be some morphological correspondence between the radio and infrared features. Fig.~\ref{zoom.plateau} is a difference image intended to isolate the strong 8~$\mu$m features from the infrared continuum. It was formed by subtracting the 3.6 $\mu$m GLIMPSE image (not shown) from the 8~$\mu$m image (Figs.~\ref{fig.ir}(a) and \ref{fig.ir}(b)). The portion of the plateau region labelled `Z' in Fig.~\ref{fig.ir}(a) is displayed in Fig.~\ref{zoom.plateau}. The 3 and 5$\sigma$ 1.4 GHz radio contours appear to be roughly outlined by the 8~$\mu$m emission along an arc which we have highlighted with arrows in Fig.~\ref{zoom.plateau}.

\subsubsection{Wide-Field Infrared Survey Explorer}
The Wide-Field Infrared Survey Explorer \citep[WISE; ][]{wright:2010} satellite has completed observations of the first 23,600 deg$^2$ of an all-sky survey. The satellite collects images in four infrared channels centred at 3.4, 4.6, 12, and 22 $\mu$m. These channels, respectively, have angular resolutions of 6.1, 6.4, 6.5, and 12 arcsec. We obtained images from all four channels from the NASA/IPAC Infrared Science Archive\footnote{\scriptsize\url{http://irsa.ipac.caltech.edu/applications/wise}}. 
\par 
It is not a straightforward to convert from the `data unit' amplitude of the WISE preliminary release images to a surface brightness scale \citep{wright:2010}. For the purpose of comparing the morphology of features between GLIMPSE and WISE images, we treat the WISE images as if they have an approximately linear, arbitrary surface brightness scale. After accounting for the lower resolution, the 12 $\mu$m WISE image shown in Fig.~\ref{fig.ir}(c) has essentially the same features as the 8 $\mu$m image in Figs.~\ref{fig.ir}(a) and \ref{fig.ir}(b). Surprisingly, the 22~$\mu$m WISE image (Fig.~\ref{fig.ir}(d)) shows that the filaments persist at longer wavelengths where emission from PAHs is not expected \citep[and references therein]{tielens:2008}. Fig.~\ref{fig.ir}(d) also highlights the several arcmin nebula which overlaps the southern filament, where strong PAH emission exhibits several sharp changes in direction.
\par 
The filaments appear to be dominated by PAHs in 3.6 to 8~$\mu$m regime. We consider the persistence of these features at 22~$\mu$m in section~\ref{discussion}. The nebula identified at 22~$\mu$m is coincident with the radio bulge of G296.7-0.9, the strongest PAH emission, and the locations of several previous radio measurements \citep[and references therein]{manchester:1969,caswell:1987,kuchar:1997}.

\subsection{H$\alpha$}
\subsubsection{United Kingdom Schmidt Telescope}
The SuperCOSMOS H$\alpha$ Survey \citep[SHS; ][]{parker:2005} used the United Kingdom Schmidt Telescope of the Anglo-Australian Observatory to observe nearly 4000 deg$^2$ of the Galactic plane in the H$\alpha$ line and with an adjacent continuum filter. These images have an angular resolution $\sim$ 1 arcsec. We have obtained the H$\alpha$ and continuum images of G296.7-0.9 from the data archive\footnote{\scriptsize\url{http://www-wfau.roe.ac.uk/sss/halpha}}. 
\par
\begin{figure}
\begin{center}
\includegraphics[width=0.45\textwidth,angle=-90]{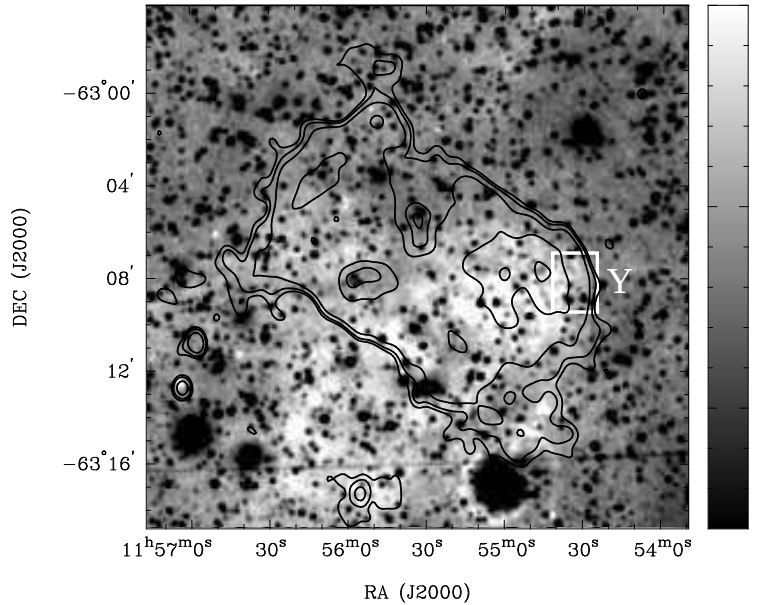}
\caption{Smoothed H$\alpha$ image of G296.7-0.9 from the SHS. This image was formed by taking the ratio of the H$\alpha$ image to the adjacent continuum, then smoothing to a resolution of 10 arcsec. The overlaid contours are the 3, 5, 10, 30, and 50$\sigma$ radio contours of Fig.~\ref{fig.radio}(b). Note the reversal of the grayscale transfer function compared to Figs.~\ref{fig.radio}-\ref{zoom.plateau}. The region labelled `Y' is shown in Fig.~\ref{zoom.halpha}. Note that the line passing from $\delta_{2000}=-63:16:15$ through the 3$\sigma$ radio contour is an image artefact.}\label{halpha}
\end{center}
\end{figure}

 
To isolate the H$\alpha$ spectral line, we have divided the H$\alpha$ image by its associated continuum image.  The resulting quotient image, smoothed to a resolution of 10 arcsec, is shown in Fig.~\ref{halpha}. H$\alpha$ emission is coincident with most of the extent of G296.7-0.9 and extends well beyond the 1.4 GHz boundary to the south of the source \citep{russeil:1997}. There is also a low surface brightness band of emission that enters the field from the eastern edge of the image, passes through the southeastern arm, and exits the field in the north. There is a sharp falloff of H$\alpha$ near the northwestern radio boundary.
\par
\begin{figure}
\begin{center}
\includegraphics[width=0.45\textwidth,angle=-90]{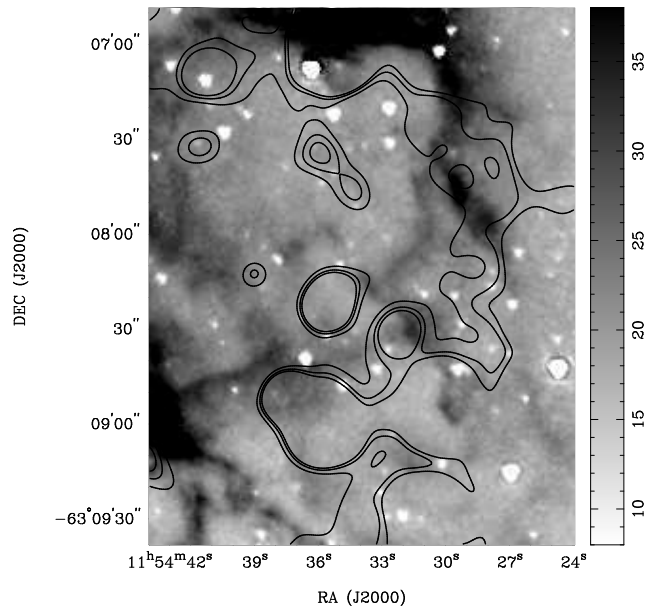}
\caption{Continuum-subtracted 8 $\mu$m GLIMPSE image of `Y' region in Figs.~\ref{fig.ir}(a) and \ref{halpha}. This image was formed by subtracting the 3.6 $\mu$m image (not shown) from the 8 $\mu$m image in order to isolate the strong 8 $\mu$m features. The scale is MJy sr$^{-1}$ and the resolution is 1.2 arcsec. The H$\alpha$ contours are arbitrary levels of Fig.~\ref{halpha}, which have a resolution of 10 arcsec.}\label{zoom.halpha}
\end{center}
\end{figure}

 
Fig.~\ref{zoom.halpha} is a difference image between the 8 $\mu$m and 3.6 $\mu$m images with contours of H$\alpha$ intensity from Fig.~\ref{halpha}. The image is centred on the region labelled `Y' in Figs.~\ref{fig.ir}(a) and \ref{halpha}. The three arbitrary contours have a similar radius of curvature. If these accurately represent the H$\alpha$ emission in the region, the H$\alpha$ and the bright 8 $\mu$m emission may have a similar morphology over several arcminutes.

\section{Discussion}\label{discussion} 
A single strong shock with negligible cosmic ray pressure propagating through a magnetised medium is expected to emit linearly polarised radio synchrotron radiation with a power-law spectral index of $\alpha$=-0.5. The emission from G296.7-0.9 is observed to be linearly polarised in sparse patches up to about 5\% of the total intensity, whereas SNRs are typically polarised in the 10-20\% range \citep[e.g.,][]{dickel:1991,reynolds:1993}. Since shock acceleration is the only known mechanism for diffuse linearly polarised synchrotron emission, any level of diffuse linear polarisation identifies a region of shock-accelerated charged particles. The soft thermal X-rays coincident with the eastern lobe of the bilateral source and the particularly bright X-ray arc along the southern rim are suggestive of shock heating of swept-up ISM or hot ejecta. Given the spectral index, the linearly polarised shell-like radio morphology and soft thermal X-rays, G296.7-0.9 is clearly a SNR.
\par
Several open questions can be addressed with the data presented here. We further discuss the the remnant's environment, age and properties, and its morphology.

\subsection{Environment}\label{disc:environment}
The SST spectra and image presented in fig. 1 of \cite{rho:2011} show that the sharply bounded southern portion of SNR N132D is lined by significant amounts of dust, identified by strong emission in the longer wavelength SST bands. The relative intensity of filaments to point-like sources at 12 $\mu$m (Fig.~\ref{fig.ir}(c)), where PAH out-of-plane bending modes are expected to radiate strongly \citep[and references therein]{tielens:2008} support the conclusion that the filaments are PAHs. However, PAHs should not substantially contribute to the 22~$\mu$m image (Fig.~\ref{fig.ir}(d)), where the filaments are still visible. If the 22 $\mu$m WISE data can be used as a stand-in for SST observations above 20 $\mu$m as used by \cite{rho:2011}, Fig.~\ref{fig.ir}(d) suggests that the infrared filaments may contain some cool dust or thermal gas in addition to PAHs. Moreover, this would suggest that the 22 $\mu$m nebula coincident with strong PAH emission in western limb of SNR G296.7-0.9 is dominated either by cool dust or thermal emission. 
\par
The single-dish Parkes telescope is sensitive to much larger scales than the ATCA and MOST and averages all emission in its $\sim$ arcminute field-of-view. Given the co-location of the pointing centre of the Parkes thermal spectrum -- the white `+' in Fig.~\ref{fig.radio}(a), radio recombination lines, 22 $\mu$m nebula, diffuse H$\alpha$ emission, and surrounding PAHs, it seems very likely that the \hii region G296.593-0.975 of \cite{caswell:1987} and SNR G296.7-0.9 are separate sources, but proximally located, if not interacting. 
\par
\cite{caswell:1987} measure a velocity for G296.593-0.975 of +25 km/s, relative to the local standard of rest. Using relevant rotation curves \citep{brand:1993,mcclure-griffiths:2007}, R$_0$=8.5 kpc, and $\Theta_0$=220 km s$^{-1}$, the \hii region has a distance of 9 $\pm$ 2 kpc. In subsequent calculations, we assume a distance of 9$d_9$ kpc to SNR G296.7-0.9 as well, where $d_9$ is a scaling factor. The linear dimensions for SNR G296.7-0.9 are then $\sim 37d_9  \times 22d_9$ pc$^2$. 

\subsection{Age \& Properties}\label{disc:age}
We do not observe a steep spectral index, excessive linear polarisation, or other indications that SNR G296.7-0.9 has reached the radiative stage of evolution \citep[and references therein]{green:1988,sutherland:2003}.
\par
For a lower age limit, we assume the remnant was recently expanding freely. Core-collapse supernovae generally produce higher shock velocities than the thermonuclear variety, so we use the typical initial core-collapse shock break-out velocity of $\sim$ 5000 km/s \citep{reynolds:2008} to place a lower bound of $\sim$ (2.8 $\pm$ 0.6)$d_9$ kyrs on the age of the remnant. 
\par
The Sedov-Taylor limit provides an upper bound on the age. Assuming that the X-ray emitting mass is completely ionised and that the ion and electron temperatures are equal, the shock velocity can be inferred from the X-ray temperature, $16 kT/(1.83 m_{\rm{p}})\approx V^2$, where $m_{\rm{p}}$ is the proton mass, $k$ is Boltzmann's constant, and $V$ is the shock velocity. We infer a shock velocity of $V=300 \pm 10$ km/s from the MEKAL temperature of 110 eV. For an adiabatic SNR,
\begin{center}
\begin{equation}
R=1.15\left(\frac{E_0 t^2}{n_0 m}\right)^{0.2}\label{radius}
\end{equation}
\end{center}

and thus,

\begin{center}
\begin{equation}
\dot{R}=V=2R/5t
\end{equation}
\end{center}

where E$_0$ is the explosion energy, t is the age, $\rm{m}$ is the expectation value of the mass per particle, and $\rm{n_0}$ is the ambient density. From the shock velocity and reduced diameter, we find that SNR G296.7-0.9 has a maximum age of (19 $\pm$ 4)$d_9$ kyrs.
\par

\subsection{Progenitor}
We expect that most SNRs result from the collapse of a massive progenitor near the Galactic plane  \citep{tammann:1994,mckee:1997}. Association with a neutron star would definitively identify SNR G296.7-0.9 remnant as resulting from the core-collapse mechanism. 
\par 
We have conducted a search for a pulsar wind nebula or pulsar by imaging just the longest five baselines of the 1.4 GHz ATCA data from both the 6B and 750A configurations. The images that results from using uniform weighting and the CLEAN deconvolution algorithm has a resolution of 6.7 $\times$ 4.4 arcsec$^2$ and the noise level of the total intensity image is 0.18 \mjy . An image of linear polarisation was formed by making linear polarisation images for each of the 12 channels, then combining them in an error-weighted average.  
\par
For a transverse velocity $< 10^3$ km/s, we would expect a compact stellar remnant to be within 8T$_{19}$d$_9^{-1}$ arcmin of its birthplace, where T$_{19}$ is a scaling factor for the true age relative to 19 kyrs. There are three sources within 8 arcmin of the inferred centre of the remnant that are detected above 5$\sigma$ (0.90 \mjy ). The positions, primary beam corrected flux, and distance from the remnant's centre are presented in Table~\ref{psrs}. These sources are unresolved and exhibit no linear or circular polarisation, nor are they identifiable in X-rays. We conclude that there is no evidence for a compact stellar remnant above a limit of 0.9 \mjy . 

\begin{center}
\begin{table}
\caption{Radio sources within 8 arcmin of the centre of SNR G296.7-0.9: location, 1.4 GHz flux, and angular distance from the centre of the remnant.}\label{psrs}
\begin{tabular}{cccc}
\hline $\alpha_{2000}$ & $\delta_{2000}$ & S$_{1.4\rm{ GHz}}$ & Distance\\
\hline 
\vspace{2mm} 11:55:06.7 & -63:07:46 & 1.0 $\pm$ 0.2 mJy & 3.3 arcmin\\
\vspace{2mm} 11:55:54.4 & -63:13:04 & 1.9 $\pm$ 0.3 mJy & 6.7 arcmin\\
\vspace{2mm} 11:56:11.3 & -63:08:26 & 1.1 $\pm$ 0.4 mJy & 4.5 arcmin\\
\hline
\end{tabular}
\end{table}
\end{center}

\subsection{Morphology}
SNR G296.7-0.9 is shell-like with bright, approximately diametric limbs. Narrow elongated radio structures similar to those seen in G46.8+0.3 \citep[HC30; ][]{dubner:1996} cross the bilateral axis between limbs, and distinct cavities of low radio surface brightness protrude into the western limb in Fig.~\ref{fig.radio}(b).
\subsubsection{The Shell}\label{disc:shell}
With our data, we can consider two potential mechanisms for the morphology of SNR G296.7-0.9: shock-compressed ambient magnetic fields \citep[e.g.,][]{vanderlaan:1962,whiteoak:1968}, and an environment with ambient density or magnetic field gradients \citep{orlando:2007}.
\par 
Lines of constant Galactic latitude are inclined at an angle of 103 deg North through East, with respect to the J2000 coordinate system shown here. The bilateral axis therefore makes an angle of 40 $\pm$ 20 deg with respect to the Galactic plane, which puts SNR G296.7-0.9 near the median of orientation angles found for a set of carefully  selected bilateral supernova remnants \citep[fig. 12 of][]{gaensler:1998a}. The remnant is a sizeable distance, $\sim$140d$_9$ pc, from the Galactic plane, where the magnetic field may not be parallel to lines of constant Galactic latitude. If the local magnetic field is not aligned with the plane of the Galaxy, the compression of ambient magnetic fields could be responsible for the bright limbs of the remnant. However, since narrow elongated structures cross the axis, the magnetic interpretation is either not responsible for the bilateral morphology, or it is only one of the effects at work in SNR G296.7-0.9.
\par
The simulations of \cite{orlando:2007} with a magnetic field directed along the bilateral axis and a gradient in magnetic field strength or density step along the line-of-sight reproduce bright limbs and a central filament similar to the radio morphology of SNR G296.7-0.9. If SNR G296.7-0.9 is either embedded in or adjacent to an \hii region, a gradient or step in ambient density along the line-of-sight is plausible; additionally, the bulge-like enhancement of radio surface brightness in the western limb could be a difference in integrated column of radio emission along the two directions.  This scenario might also contribute to the lack of X-rays in the western lobe, provided the initial SNR-cloud interaction began substantially long ago \citep{mac-low:1994,orlando:2006}. On the other hand, and similar to ambient field compression case, a density or magnetic field gradient does not explain the southernmost narrow elongated structure. 
\par
We note that the directions of the ambient field inferred from the magnetic compression mechanism and from simulations of \cite{orlando:2007} are the same. In these interpretations, the projection of the Galactic field on the sky aligns with the bilateral axis of the remnant.

\subsubsection{The Plateau and Arms}
Extensions of low surface brightness protruding near the bright arcs of bilateral remnants are observed in several other sources, including 3C 391, G298.6-0.0, and G67.7+1.8 \citep{reynolds:1993,whiteoak:1996,kothes:2006}. In the case of SNR G296.7-0.9, we consider three possible causes for these faint features: thermal radio emission, a radio halo, and the breakout of the shock into a lower density cavity.
\par
The arms and plateau are not well-detected enough to measure their spectral index. However, since there are no local excesses of H$\alpha$ nor 22~$\mu$m emission and they are connected to bright non-thermal emission above the noise levels in both Figs.~\ref{fig.radio}(b) and \ref{fig.radio}(c), we see the possibility that they are thermal as unlikely.
\par 
Radio haloes due to the streaming of relativistic charged particles ahead of a shock front were postulated by \cite{achterberg:1994}. The electron diffusion length is, $L=\lambda c/3V$ \citep{moffett:1994b}, where $\lambda$ is the charged particle mean-free path (MFP) and $V$ is the shock speed. Assuming that the shock speed is roughly the same everywhere along its edge, differences in diffusion length between regions result from differences in the average energy of the charged particle distribution, the strength of the magnetic field, or the local amplitude of Alfv\'{e}n waves,
\begin{center}
\begin{equation}
\frac{L_1}{L_2} \sim \frac{\gamma_1 B_2}{\gamma_2 B_1}\left( \frac{\delta B_2 B_1\cos{\theta_{1}}}{B_2 \delta B_1  \cos{\theta_{2}}} \right)^2
\end{equation}
\end{center}
where the numerical subscripts indicate different regions, $\theta$ is the angle between the shock normal and the upstream magnetic field, $B$, ($\delta B/B)^2$ is the amplitude of upstream Alfv\'{e}n waves, and $\gamma$ is the average Lorentz factor of the emitting electrons. 
\par
After creating a profile of the surface brightness at select sites, we measure the diffusion length using the method of \cite{moffett:1994a} -- by defining the shock front as the inflection point of the profile and estimating the distance to the edge of the remnant, defined by the 3$\sigma$ contour. The shortest diffusion lengths, typically at the northwestern and southeastern rim are unresolved. We use the 10$\sigma$ contour to define the inner boundary of the diffusion length in regions where an inflection point is not obvious; e.g., the arm and plateau regions. Given the resolution of the image presented in Fig.~\ref{fig.radio}(b), we are able to identify $\frac{L_1}{L_2}$ of up to $\sim$6.
\begin{figure}
\begin{center}
\includegraphics[scale=0.45, angle=0]{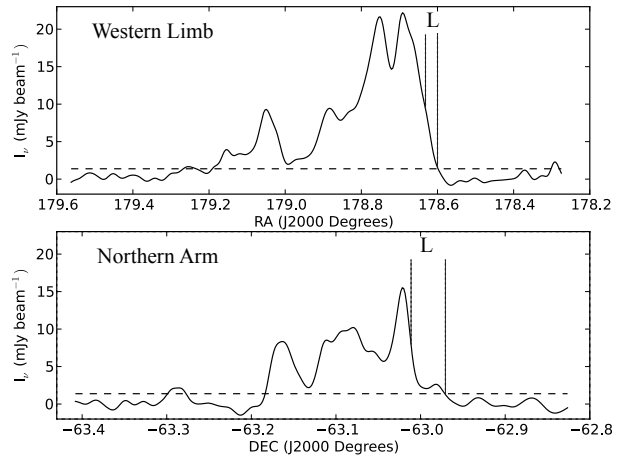}
\caption{1.4 GHz surface brightness profiles of G296.7-0.9 from Fig.~\ref{fig.radio}(b). The horizontal dashed line is the 3$\sigma$ level from Fig.~\ref{fig.radio}(b) and the vertical dotted lines show the relative positions of the inflection point and edge of the remnant, the distance between which we use as an estimate of the diffusion length. The top panel is the surface brightness profile of a horizontal slice through Fig.~\ref{fig.radio}(b) at $\delta_{2000}=$ -63:08:22 and the bottom panel is a vertical slice through $\alpha_{2000}=$ 11:55:49.}\label{profiles}
\end{center}
\end{figure}
\par
Unless the distance to SNR G296.7-0.9 is significantly less than 9 kpc, a radio halo of such a large extent seems unlikely given the MFP derived for other remnants; e.g., scaling the findings of \cite{achterberg:1994} to a distance of 9 kpc, we would expect $\lambda \le 0.07d_9^{-1}$ arcsec. The magnetic field is not likely to be reduced by greater than a factor of two from one synthesised beam to a neighbouring beam. Therefore, the final possibility for a radio halo lies in the $\theta$-dependence of the diffusion length. If the large-scale ambient field is uniform, it is difficult to envision an orientation of the remnant that would allow particle streaming only in the arms and plateau regions.
\par
If the plateau is due to the break-out of a shock into a lower density medium, the lack of an inflection point in the surface brightness profile suggests that cavities in the western limb may be related to the plateau. \cite{fulbright:1990} and \cite{moffett:1994a} suggest that by assuming a linear gradient of the field from one side of a remnant to the other, the ratio of radio surface brightness between two limbs is, $\zeta$ $ \sim \left(B_2/B_1\right)^{1.2}$. Furthermore, under the assumption that synchrotron emitting cosmic rays are accelerated to a constant fraction of post-shock pressure, a gradient in electron density implies that $\zeta$ $ \sim \left(n_2/n_1\right)^{1.8}$ \citep{reynolds:1981}. To examine the possibility that the cavities are due to variations in magnetic field strength or shock break-out, we use these $\zeta$ relations under the assumption that they hold for smaller scales than  \cite{fulbright:1990} and \cite{moffett:1994a} proposed. 
\par
The ratio of surface brightness between the cavity and its radial or circumferential boundary gives $\zeta$ $\sim$ 1.6 $\pm$ 0.3. Therefore, either the magnetic fields or density change by about a factor of two from one synthesised beam to the next. If the limb brightening of bilateral remnants is due to compression of magnetic fields from the interstellar medium, which requires the fields to be well-ordered and have a scale length at least as large as the length of the bright limb, it would be difficult to attribute the small, sharp boundaries of the cavities to a local variation in the magnetic field. On the other hand, the breakout of a shock from a high density region to one of low density could easily provide a factor of two in density and the sharp boundary. Moreover, the shock breakout conclusion is supported by the low levels of H$\alpha$ emission in the plateau region, and by the boundary of the cavities closest to the centre of the remnant being lined by 8 $\mu$m emission. 
\par
Given the radio surface brightness profiles, H$\alpha$, and infared properties, we propose that the cavities and plateau are possibly related by the breakout of the SNR shock into a low density medium. In this interpretation, the shock would have swept from a dense region with a non-smooth boundary between the 10 and 30$\sigma$ contours of Fig.~\ref{fig.radio}(b), into a very low density region. The shock speed would increase in the region of lower density and in time, this higher speed could allow the shock to form the faint plateau and arms of radio emission that are observed.

\section{Conclusions}
Based upon the non-thermal spectral index, existence of linear polarisation, and coincidence with an arc of thermal X-rays, we confirm that  G296.7-0.9 is a supernova remnant. 
\par
A thermal single-dish radio spectrum, radio recombination lines and 22 $\mu$m emission, in concert with morphological similarities between emission in different wavebands suggest that this remnant may be co-located with an \hii region at a distance of $9 \pm 2$  kpc. We suggest that the most plausible explanation for the arm-like extensions, plateau, and low surface brightness cavities is the breakout of the SNR shock from a circumstellar distribution of matter into a lower density cavity in the ISM. Two feasible mechanisms for the bilateral morphology of SNR G296.7-0.9 suggest that the large-scale ambient magnetic field points along the SNR's axis of symmetry.  We see no evidence for an associated neutron star or pulsar wind nebula. 
\par
Further study of this remnant would benefit from a long integration, high spatial and spectral resolution X-ray observation. In particular, modern telescopes could easily perform a search for a pulsar, clarify the extent of the X-ray source, and place tighter constraints on the mechanism of the X-ray emission. Higher frequency radio observations may also be useful to examine the polarisation properties of the remnant. Furthermore, infrared spectra could provide a more complete understanding of the filaments and nebula that appear to be associated with SNR G296.7-0.9.  

\section*{Acknowledgments} 
WJR is grateful to the School of Physics at The University of Sydney for financial support through an International Denison Postgraduate Award. We thank Gemma Anderson, Martin Cohen, Lisa Harvey-Smith, Greg Madsen, Naomi McClure-Griffiths, and Tim Robishaw for useful discussions. We also acknowledge our anonymous referee for their helpful comments.
\par
The Australia Telescope is funded by the Commonwealth of Australia for operation as a National Facility managed by CSIRO. This research has made use of data and software provided by the High Energy Astrophysics Science Archive Research Center (HEASARC), which is a service of the Astrophysics Science Division at NASA/GSFC and the High Energy Astrophysics Division of the Smithsonian Astrophysical Observatory. This work has also used data from the NASA/IPAC Infrared Science Archive, which is operated by the Jet Propulsion Laboratory, California Institute of Technology, under contract with the NASA. H$\alpha$ and optical contiuum images were taken from the SuperCOSMOS Science Archive, prepared and hosted by the Wide Field Astronomy Unit, Institute for Astronomy, University of Edinburgh, which is funded by the UK Science and Technology Facilities Council.

\def\aj{AJ}%
\def\actaa{Acta Astron.}%
\def\araa{ARA\&A}%
\def\apj{ApJ}%
\def\apjl{ApJ}%
\def\apjs{ApJS}%
\def\ao{Appl.~Opt.}%
\def\apss{Ap\&SS}%
\def\aap{A\&A}%
\def\aapr{A\&A~Rev.}%
\def\aaps{A\&AS}%
\def\azh{AZh}%
\def\baas{BAAS}%
\def\bac{Bull. astr. Inst. Czechosl.}%
\def\caa{Chinese Astron. Astrophys.}%
\def\cjaa{Chinese J. Astron. Astrophys.}%
\def\icarus{Icarus}%
\def\jcap{J. Cosmology Astropart. Phys.}%
\def\jrasc{JRASC}%
\def\mnras{MNRAS}%
\def\memras{MmRAS}%
\def\na{New A}%
\def\nar{New A Rev.}%
\def\pasa{PASA}%
\def\pra{Phys.~Rev.~A}%
\def\prb{Phys.~Rev.~B}%
\def\prc{Phys.~Rev.~C}%
\def\prd{Phys.~Rev.~D}%
\def\pre{Phys.~Rev.~E}%
\def\prl{Phys.~Rev.~Lett.}%
\def\pasp{PASP}%
\def\pasj{PASJ}%
\def\qjras{QJRAS}%
\def\rmxaa{Rev. Mexicana Astron. Astrofis.}%
\def\skytel{S\&T}%
\def\solphys{Sol.~Phys.}%
\def\sovast{Soviet~Ast.}%
\def\ssr{Space~Sci.~Rev.}%
\def\zap{ZAp}%
\def\nat{Nature}%
\def\iaucirc{IAU~Circ.}%
\def\aplett{Astrophys.~Lett.}%
\def\apspr{Astrophys.~Space~Phys.~Res.}%
\def\bain{Bull.~Astron.~Inst.~Netherlands}%
\def\fcp{Fund.~Cosmic~Phys.}%
\def\gca{Geochim.~Cosmochim.~Acta}%
\def\grl{Geophys.~Res.~Lett.}%
\def\jcp{J.~Chem.~Phys.}%
\def\jgr{J.~Geophys.~Res.}%
\def\jqsrt{J.~Quant.~Spec.~Radiat.~Transf.}%
\def\memsai{Mem.~Soc.~Astron.~Italiana}%
\def\nphysa{Nucl.~Phys.~A}%
\def\physrep{Phys.~Rep.}%
\def\physscr{Phys.~Scr}%
\def\planss{Planet.~Space~Sci.}%
\def\procspie{Proc.~SPIE}%
\newpage 
\bibliographystyle{mn2e}
\bibliography{refs}

\end{document}